\documentstyle[12pt]{article}
\begin{document}

\begin{titlepage}
\title{Quintessence duality}

\author{{R. de Ritis$^{1,2}$, A. A. Marino$^{2,3}$, C. Rubano$^{1,2}$, P. Scudellaro$^{1,2}$} \\ 
{\em \small $^1$Dipartimento di Scienze Fisiche, Universit\`{a} di 
Napoli,} \\
{\em \small Complesso Universitario di Monte S. Angelo, Via Cintia, Ed. N, 80126 Napoli, Italy;} \\
{\em \small $^2$Istituto Nazionale di Fisica Nucleare, Sezione di Napoli,} \\
{\em \small Complesso Universitario di Monte S. Angelo, Via Cintia, Ed. G, 80126 Napoli, Italy;} \\
{\em \small $^3$Osservatorio Astronomico di Capodimonte,} \\ 
{\em \small Via Moiariello, 16, 80131 Napoli, Italy.}}
	      \date{}
	      \maketitle
	      \begin{abstract}
We join quintessence cosmological scenarios with the
duality simmetry existing in string dilaton cosmologies. Actually, we
consider the tracker potential type $V = V_0/{\varphi}^{\alpha}$ and show
that duality is only established if $\alpha =  - 2$.
               \end{abstract}

\

\noindent PACS number(s): 98.80.Cq, 98.80.-k, 98.80.Hw, 04.20.Jb

\

\noindent Corresponding author: A.A. Marino, Osservatorio Astronomico di Capodimonte, Via Moiariello, 16, 80131 Napoli, Italy \\
tel. +39 081 5575547, fax +39 081 456710 \\
e-mail marino@na.astro.it \\
	      \vfill
	      \end{titlepage}

In this short letter we show how duality simmetry can be connected with the
so\,-\,called quintessence theories.

Duality (see the recent review paper on this topic by M. Gasperini
\cite{Gasp99}
and the almost complete references therein) is an extra\,-\,simmetry existing
(in the cosmological arena\,: we will use a standard 3+1
Friedmann\,-\,Robertson\,-\,Walker splitting of spacetime, i.e. the
cosmological principle) in the minisuperspace associate with
scalar\,-\,theories of gravity (see \cite{CdRRS96} and the references
therein). Essentially this simmetry is such that if $a(t)$ is a cosmological
solution, then $a^{-1}(t)$ is also a solution when the scalar field (dilaton)
present in this approach to gravity undergoes to the trasformation $\varphi
\rightarrow \bar{\varphi} = \varphi - (3/2) \ln{a}$. Of course this
simmetry implies the existence of a new fundamental constant $\lambda_S$, i.e.
the fundamental string scaling. The reason of the great interest in string
theory is very well described in the quoted paper by Gasperini \cite{Gasp99}
and we make reference to it for a discussion on this topic.

The second ingredient we discuss in our paper is quintessence, which
has been developed recently \cite{CDS}, \cite{Ostrik}, \cite{Perl3},
\cite{Stein}, \cite{SWZ}, \cite{ZWS} because the astronomical observations
suggest that there is present in the Universe a very large amount of energy
density with a negative pressure. We may consider it as given by a scalar
field $\varphi$ slowly rolling down its potential and such that $-1 <
w_{\varphi} < 0$, where $w_{\varphi} = p_{\varphi}/\rho_{\varphi}$, being
$p_{\varphi}$ and $\rho_{\varphi}$ the pressure and the energy density of the
scalar field. Actually recent considerations \cite{Perl3} fix this interval
to be smaller\,:\,$-1 < w_{\varphi} \le - 0.6$.
\footnote{We like to remember that a cosmological constant $\Lambda$ which
can be connected with vacuum energy density has $w_{\Lambda} = -1$.}
As a matter of fact, quintessence has been advanced resuming, generalizing
and suitably readapting older ideas on cosmology with a scalar field, i.e.
scalar\,-\,tensor theories of gravity. (See also the introduction of the
so\,-\,called x\,-\,matter \cite{Turner}, \cite{Chiba1}, \cite{Chiba2},
\cite{Chiba3}.)
One of the most common potential used in quintessence theories is of the type
$V = V_0 {\varphi}^{- \alpha}$, (where $V_0$ and $\alpha > 0$ are parameters);
its discussion can be found in the literature already quoted, but see also
\cite{Paolo} because we are going to analyse quintessence in a
non\,-\,minimal coupled theory.

These are the two ingredients we are going to join in what follows.

We will refer to the general string action\,:

\begin{equation}
s = \frac{1}{\lambda_{s}^{d-1}} \int{d^{d+1}x \sqrt{g} e^{-2\varphi}
[ R + 4 (\nabla \varphi )^2 + V(\varphi) ] } +
\int{d^{d+1}x \sqrt{g} L_M} \,,
\label{eq: uno}
\end{equation}
where $L_M$ rapresents all the others matter source, and the antisymmetric
tensor $B_{\mu \nu}$ present in the string theory has been taken zero as
usual in string cosmology; finally, we will assume $d = 3$.

Let us consider cosmological non\,-\,minimally coupled models described by
the general action\,:

\begin{equation}
{\cal A} = \int{d^4 x \sqrt{g} \left \{ \left [ F(\varphi)R + \frac{\gamma}
{2} (\nabla \varphi )^2 -  V(\varphi) \right ] + L_M \right \} } \,.
\label{eq: due}
\end{equation}
Here, the coupling function $F = F(\varphi)$ is completely generic, as well
as, at this level, the potential $V(\varphi)$. We suppose that in (\ref{eq:
uno}) and (\ref{eq: due}) $L_M = 0$, that is, we consider a situation in
which there are no sources other than the scalar field. Furthermore we have
introduced a generic coefficient $\gamma$ in the kinetic part because of
the discrepancy existing, on this coefficient, between (\ref{eq: uno}) and
the standard non\,-\,minimal coupled scalar\,-\,tensor action. From
(\ref{eq: due}), in the FLRW cosmological case, we get the following
`point\,-\,like' Lagrangian\ defined in the minisuperspace $a(t),
\varphi(t)$:

\begin{equation}
{\cal L} = a^3 \left [
6 F(\varphi) \left ( \frac{\dot{a}}{a} \right )^2
+ 6 F'(\varphi) \dot{\varphi} \left ( \frac{\dot{a}}{a} \right )
+ \frac{\gamma}{2} \dot{\varphi}^2 - V(\varphi) \right ] \ .
\label{eq: tre}
\end{equation}
Let us choose the potential commonly used in the quintessence approach,
i.e.\,:

\begin{equation}
V = V_0 \varphi^{\alpha} \ ,
\label{eq: quattro}
\end{equation}
considering now $\alpha$ completely arbitrary
\footnote{We do not enter in details on the way of putting constraints on this
parameter because the goal of our paper is different; in any case see
\cite{Leib}.}.
Let us change $\varphi \rightarrow \varphi_0 e^{\beta
\psi}$ and write the coupling as\,:
\begin{equation}
F = F_0 e^{2 \beta \psi} f(\psi) \ ,
\label{eq: cinque}
\end{equation}
being $(F_0, \varphi_0, \beta)$ constants and $f(\psi)$ a generic function
of the new $\psi$\,-\,field. Using these new variables and functions,
Lagrangian (\ref{eq: tre}) becomes\,:

\begin{equation}
 {\cal L} = a^3 e^{2 \beta \psi} \left [
6 F_0 f \left ( \frac{\dot{a}}{a} \right )^2 + 6 F_0 (2 \beta f + f')
\dot{\psi} \left ( \frac{\dot{a}}{a} \right ) + \frac{\gamma}{2}
{\varphi_0}^2 \beta^2 \dot{\psi}^2
-\frac{V}{\varphi_0}^{\alpha} e^{-(2\beta + \alpha)\psi} \right ] \ .
\label{eq: trep}
\end{equation}
(The prime indicates derivation with respect to $\psi$.)

Let us now perform the (standard) transformation :
\begin{equation}
\psi = \psi (a, \phi) = k \phi + \frac{3}{2} \ln{a} \ ,
\label{eq: sei}
\end{equation}
where $k$ is a new parameter. The piece
\begin{displaymath}
a^3 e^{2 \beta \psi} = a^3 e^{2 \beta k \phi} e^{\beta \ln{a^3}} = a^3 a^{3
\beta} e^{(2 k \beta) \phi} \ ,
\end{displaymath}
if $\beta = -1$, becomes $a^{-3} a^3 e^{- 2 k \phi} = e^{-2 k \phi}$.
Substituting (\ref{eq: sei}) in (\ref{eq: trep}) gives\,:
\begin{displaymath}
{\cal L} = e^{ -2 k \phi} \left [
\left ( \frac{\dot{a}}{a} \right )^2
\left (- 12 F_0 f + 9 F_0 f' \frac{9}{8} \gamma \varphi_0^2 \right )
+ k \left ( -12 F_0 f + 6 F_0 f' \right. \right.
\end{displaymath}
\begin{equation}
\ \ \left. \left. + \frac{3}{2} \gamma \varphi_0^2 \right )
\dot{\phi} \left ( \frac{\dot{a}}{a} \right ) \right ] + e^{ -2 k \phi}
\left [ \frac{\gamma}{2} \varphi_0^2 k^2 \dot{\phi}^2 - V_0 \varphi_0^{\alpha}
- e^{\psi (2 - \alpha)} \right ] \ .
\label{eq: sette}
\end{equation}
Let us require that $f$ is a solution of the equation :
\begin{equation}
\frac{1}{2} f' - f + \frac{\gamma \varphi_0^2}{8 F_0} = 0 \ .
\label{eq: otto}
\end{equation}
Because of (\ref{eq: sei}) we can choose, as a particular solution of
(\ref{eq: otto}), $f = f_0$; then, we have $f_0 = \frac{\gamma
\varphi_0^2}{8 F_0}$, which substituted in the coefficient of
$(\dot{a}/a)^2$ gives\,:
\begin{displaymath}
{\cal L} = e^{-2 k \phi} \left [
- \frac{3}{2} \varphi_0^2 \left (1 - \frac{3 \gamma}{4} \right )
\left ( \frac{\dot{a}}{a} \right )^2 +
\left ( \frac{\gamma}{2} \varphi_0^2 k^2 \right ) \dot{\phi}^2 -
V_0 \varphi_0^{\alpha} e^{\psi (2 - \alpha)} \right ] \ .
\end{displaymath}
If we now suppose, for example, that $\varphi_0 = - \sqrt{8}$, $k^2 = 1$,
$\gamma = 1$, we get ($\bar{V}_0 \equiv - V_0 \varphi_0^{\alpha}$)\,:
\begin{equation}
{\cal L} = e^{{\pm} 2 \phi} \left [
-3 \left ( \frac{\dot{a}}{a} \right )^2 + 4 \dot{\phi}^2 + \bar{V}_0
e^{(2 - \alpha) \psi(a, \phi)} \right ] \ ,
\label{eq: nove}
\end{equation}
and we recover duality if $\alpha = 2$. That is, duality simmetry requires
that the potential has to be $V = V_0 \varphi_0^2$, i.e. it is not possible
to recover the quintessence standard power--law type potential.

As a final remark we want to stress that, if we compute the parameter
$\Gamma \equiv V''V/(V')^2$ using the initial potential $V = V_0
\varphi^2$, we get $\Gamma_\varphi = 1/2$, so that $\Gamma_\varphi < 1$.
\footnote{To get a tracker field, a special form of quintessence, we need to
have $\Gamma > 1$ \cite{SWZ}, \cite{ZWS}.} If we compute the same quantity
relatively to the potential $V =\bar{V}_0 e^{- 2k \phi}$ (which is the same
one used by B. Ratra \cite{Ratra}) given in (\ref{eq: nove}), after the
change (\ref{eq: sei}) of minisuperspace variables, we find $\Gamma_{\phi}
= 1$. That is, we have to carefully compute how $\Gamma$ changes under the
(more general than (\ref{eq: sei})) transformation\,:

\begin{equation}
a = a\,, \ \ \ \ \ \varphi = \varphi(a, \phi)\,.
\label{eq: dieci}
\end{equation}
A straightforward calculation shows that\,:

\begin{equation}
\tilde{\Gamma} \equiv \frac{\displaystyle{\frac{\partial^2 \tilde{V}}{\partial \phi^2}}}
{\left(\displaystyle{\frac{\partial \tilde{V}}{\partial \phi}} \right)^2}
\tilde{V}
=
\frac{\displaystyle{\frac{\partial^2 V}{\partial \varphi^2}}}
{\left(\displaystyle{\frac{\partial V}{\partial \varphi}} \right)^2} V +
\frac{1}{\displaystyle{\frac{\partial}{\partial \phi} \ln \tilde{V}}}
\left(\frac{\partial}{\partial \varphi} \frac{1}{\displaystyle{\frac{\partial \phi}
{\partial \varphi}}} \right)\,, \nonumber
\end{equation}
i.e.\,:

\begin{equation}
{\tilde{\Gamma}} = \Gamma +
\frac{1}{\displaystyle{\frac{\partial}{\partial \phi}} \ln \tilde{V}}
\left(\frac{\partial}{\partial \varphi} \frac{1}{\displaystyle{\frac{\partial \phi}
{\partial \varphi}}} \right)\,,
\label{eq: undici}
\end{equation}
where $\tilde{V} = \tilde{V}(a, \phi) = V(\varphi(\phi, a))$.

Computing the piece arising from the change (\ref{eq: sei}), and using the
exponential potential present in (\ref{eq: nove}), we get\,:

\begin{equation}
\frac{1}{\displaystyle{\frac{\partial}{\partial \phi}} \ln \tilde{V}}
\left(\frac{\partial}{\partial \varphi} \frac{1}{\displaystyle{\frac{\partial \phi}
{\partial \varphi}}} \right) = \frac{1}{2}\,, \nonumber
\end{equation}
and relation (\ref{eq: undici}) gives rise exactly to $\tilde {\Gamma} =
1$, as required by the potential present in (\ref{eq: nove}).

In a forthcoming analysis we will study relation (\ref{eq: undici}), which
naturally comes out in the string duality approach, in order to understand
whether the quintessence requirement $\Gamma > 1$ is (minisuperspace)
coordinate dependent or not.

~\\{\it Acknowledgments}.

It is a great pleasure to thank V.F. Cardone, E. Piedipalumbo, and P.
Trautmann for the discussions we had on the manuscript. This work has been
finantially sustained by the M.U.R.S.T. grant PRIN97 ``SIN.TE.SI.''.

\end{document}